\documentclass[journal=cmatex,manuscript=article]{achemso}

\usepackage[version=3]{mhchem} 
\usepackage{lineno}
\usepackage[inkscapelatex=false]{svg}

\usepackage{graphicx}

\usepackage{amsmath}
\usepackage{amssymb}
\usepackage{chemformula}
\usepackage{mhchem}
\usepackage{xr}
\usepackage{hyperref}
\DeclareUnicodeCharacter{00A0}{~}  

\graphicspath{{figs/}}

\DeclareCaptionFormat{ruleabove}{\rule{\linewidth}{0.4pt}\\#1#2#3}

\author{Fatmag\"ul Katmer}
\affiliation{Department of Chemistry, Princeton University, Princeton, NJ 08544}

\author{Milena Jovanovic}
\affiliation{Department of Chemistry, North Carolina State University, NC 27695} 

\author{Jennifer Cano}
\affiliation{Department of Physics and Astronomy, Stony Brook University, Stony Brook, New York 11794}
\altaffiliation{Center for Computational Quantum Physics, Flatiron Institute, New York, New York 10010}

\author{Lukas Muechler}
\affiliation{Department of Chemistry,The Pennsylvania State University, PA 16802}

\author{Leslie M. Schoop}
\affiliation{Department of Chemistry, Princeton University, Princeton, NJ 08544}
\email{lschoop@princeton.edu}

\title[An \textsf{achemso} demo]
  {A Recipe for Flat Bands in Pyrochlore Materials: A Chemist's Perspective}

\abbreviations{abbreviations}
\keywords{Pyrochlore, Flat Band, Bonding, Molecular Orbitals, Band Structure}

\begin{document}

\begin{abstract}

Materials in which atoms are arranged in a pyrochlore lattice have found renewed interest, as - at least theoretically - orbitals on those lattice can form flat bands. However, real materials often do not behave according to theoretical models, which is why there has been a dearth of pyrochlore materials exhibiting flat bands physics. Here we examine the conditions under which ideal ``pyrochlore bands'' can exist in real materials and how to have those close to the Fermi level. We find that the simple model used in the literature does not apply to the bands at the Fermi level in real pyrochlore materials. However, surprisingly, we find that certain oxide compounds which have oxygen orbitals inside the pyrochlore tetrahedra do exhibit near-ideal pyrochlore bands near the Fermi level. We explain this observation by a generalized tight-binding model including the oxygen orbitals. We further classify all known pyrochlore materials based on their crystal structure, band structure and chemical characteristics and propose materials to study in future experiments.
\end{abstract}

\section{Introduction}
   
 Flat-band materials are the subject of intense study in materials science\cite{Checkelsky_2024,flatbandcatalogue,Neves2024,Jovanovic2022}.  Flat bands coinciding precisely with the Fermi level provide a route to many interesting strongly correlated phenomena, such as superconductivity, magnetism or the quantum Hall effect\cite{Tian2023,Unconventional_superconductivity, Hasan2019, quaantum_hall}. However, not all flat bands give rise to the same physics. Flat bands have recently been categorized into three types, depending on where or whether the electrons localize\cite{flatbandcatalogue}:  1) Flat atomic bands, where the electrons are localized on atomic or molecular sites. These can be understood as non-interacting electrons in orbitals with non-bonding character and are not relevant for any of the exotic properties mentioned above. 2) Flat obstructed atomic bands, where the electrons are localized in covalent bonds between the atoms. 3) Flat topological bands, stemming from bonding orbitals, which cannot disperse because of destructive interference. This results in a bonding interaction with flat dispersion. The latter are the most interesting, as they can give rise to strongly correlated electrons and with emergent collective phenomena. They can be seen as delocalized electrons with no energy dispersion, resulting in strong interactions among such electrons. 
 Flat topological bands may arise from geometric frustration of the lattice. Lattice frustration causes destructive interference in electron motions, leading to the real-space localization of electron wavefunctions into so-called "compact localized states"\cite{CLS, Neves2024}, which interact across the lattice. Although the precise characterization of whether a particular flat band is topological or not is complex and depends on whether the flat band is isolated, research has focused on finding materials that host flat bands that arise through such lattice frustrations.\cite{Neves2024,Wakefield2023, CLS_Jen_1, Jen_2}
 
 Some of the very well-known lattice types that facilitate such frustration are the kagome lattice, Lieb lattice, and pyrochlore lattice\cite{Kang2020, Slot2017, Jovanovic2022, flatbandcatalogue, Neves2024}. Recently, all known compounds were cataloged by the flatness of their bands \cite{flatbandcatalogue}.  The flat band database found 6,338 flat band compounds, 1699 of which were kagome materials, 296 pyrochlore, and 721 were Lieb lattice compounds. However, a major challenge is realizing flat bands that are energetically isolated, which is regarded as a necessary condition for realizing the exciting physical phenomena described above. In particular, kagome materials have been heavily studied for potential flat bands\cite{Jovanovic2022, Kang2020, Ye2024, Kang2020_2}, but there has been limited experimental evidence for isolated kagome flat bands. This may be related to the fact that a kagome lattice is not stable by itself in a 2D form and thus will always have close-by elements to stabilize the lattice. Those close-by elements will influence the band structure and may disturb with the potential flat bands\cite{Jovanovic2022}. In this regard, pyrochlore lattices, which can be viewed as 3D versions of the kagome lattice, are more promising.

 \begin{figure}[h]
    \centering
    \resizebox{\textwidth}{!}{\includegraphics{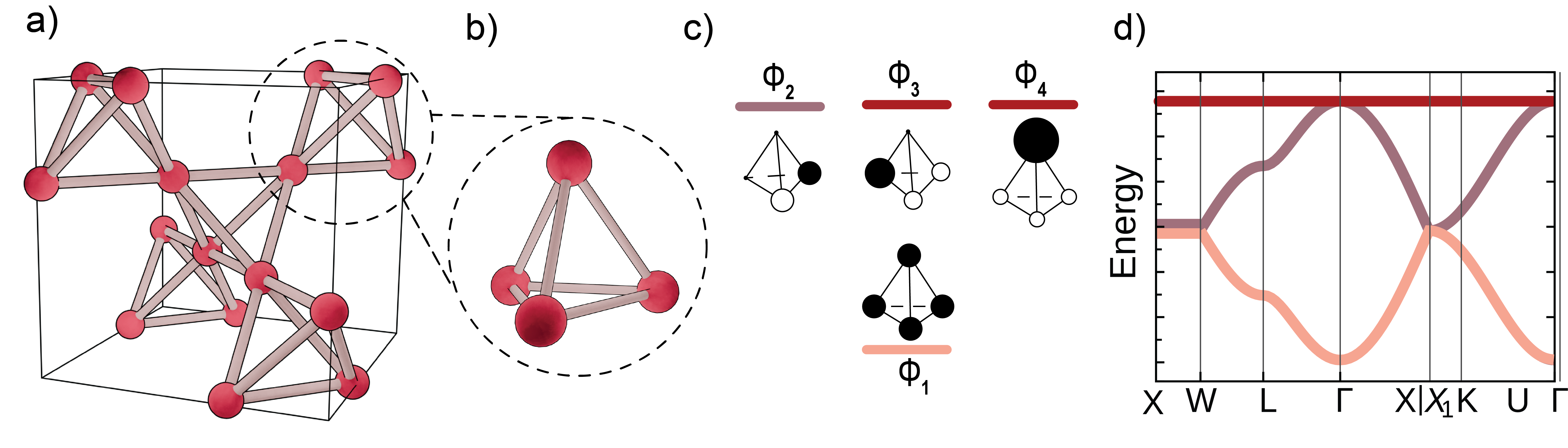}}
    \caption{Pyrochlore structure and band motif: a) The pyrochlore lattice consists of corner-shared tetrahedra. b) The primitive cell of the lattice is simple tetrahedron. c) The symmetry-adapted linear combination (SALC) of \(s\) orbitals at the \(\Gamma\) point for the primitive cell is given by:  
$\phi_1 = \psi_1 + \psi_2 + \psi_3 + \psi_4, \phi_2 = \psi_3 - \psi_4, \phi_3 = 2\psi_2 - \psi_3 - \psi_4, \phi_4 = 3\psi_1 - \psi_2 - \psi_3 - \psi_4.$ \(\phi_3\) and \(\phi_4\) (red orbitals) form the doubly degenerate flat bands, whereas \(\phi_1\) (pink orbital) and \(\phi_2\) (purple orbital) remain dispersive. d) The \textit{s} orbital tight-binding band structure yields a doubly degenerate flat band.}
    \label{intro_figure}
\end{figure}

 The pyrochlore lattice is shown in Fig. \ref{intro_figure} (a). It can be understood as a 3D lattice of corner sharing tetrahedra shown in Fig.~\ref{intro_figure} (b). 
A nearest-neighbor tight-binding model constructed from $\textit{s}$ orbitals on such a lattice yields  characteristic ``pyrochlore bands'', as shown in Fig. \ref{intro_figure} (d). The important feature is the doubly degenerate flat band at the top of the band manifold. These bands are often cited as being typical for any pyrochlore material, but a closer look at Density Functional Theory (DFT)-computed band structures of such materials shows that the mere presence of a pyrochlore lattice does not guarantee such bands. Throughout this paper, we will refer to bands that resemble Fig.~\ref{intro_figure} (d) as pyrochlore bands, and highlight them as ``clean'' if they appear without much overlap from other bands. From an orbital picture (Fig. \ref{intro_figure} (c)), we can understand the presence of the doubly degenerate flat bands by observing the symmetry-adapted linear combinations (SALCs) of the \(s\) orbitals at the \(\Gamma\) point. The linear combinations are given by the following formulas: $\phi_1 = \psi_1 + \psi_2 + \psi_3 + \psi_4$, $\phi_2 = \psi_3 - \psi_4$, $\phi_3 = 2\psi_2 - \psi_3 - \psi_4$, $\phi_4 = 3\psi_1 - \psi_2 - \psi_3 - \psi_4$. The combinations \(\phi_3\) and \(\phi_4\) form the doubly degenerate flat bands; this can be verified when altering the orbital phases which happens when the k-vector progresses. The combinations \(\phi_1\) and \(\phi_2\) remain dispersive due to their different symmetry characteristics.

Theoretical work indeed suggests the presence of flat bands in pyrochlore materials such as \ch{Pb2Sb2O7}\cite{flatbandcatalogue}. Recently, flat bands were experimentally reported in  pyrochlore material \ch{CaNi2}\cite{Wakefield2023}. Additionally, non-Fermi liquid behavior in a correlated flat band pyrochlore lattice was shown in \ch{CuV2S4}\cite{Huang2024_CuV2S4}. Moreover, in pyrochlore \ch{CeRu2} \cite{Huang2024_CeRu2} the existence of flat bands was connected to its superconducting properties. However, a comprehensive search and chemical understanding of all pyrochlore materials, differentiating when they will have pyrochlore bands or flat bands and when not, and analyzing when features of interest will be close to the Fermi level, is thus far lacking.

Here we provide an analysis of all 945 unique pyrochlore compounds that we could find in the ICSD. We find that - perhaps counterintuitively - compounds with simple chemical structures often do not have clean  pyrochlore bands at the Fermi level, whereas compounds with more complex structures do. We analyze the chemical origin of the pyrochlore bands via tight-binding models and group theory and find that bonding interactions with non-pyrochlore atoms and local symmetries provide tuning knobs to isolate flat bands in pyrochlore materials that are close to the Fermi level. 
   
Our insights will be helpful in designing future flat-band materials with different lattice types. Our main findings can be summarized as follows: Materials with pyrochlore atoms and few other atoms often do not have clean pyrochlore bands - mostly because those pyrochlore lattices are formed by \textit{p} or \textit{d} orbitals, where the multiplicity of those orbitals leads to a complex band structure. We also show that ferromagnetism can sometimes split \textit{d}-bands sufficiently to better isolate pyrochlore bands, although, also here, they are far from optimal. In contrast, pyrochlore oxides, which have much more complex structures with oxygen atoms incorporated into pyrochlore tetrahedra, may have beautiful and clean pyrochlore bands exhibiting the desired flat bands. We find that the origin here lies in the bonding interaction of the oxygen atoms with the pyrochlore orbitals and conclude that the cleanest pyrochlore bands actually arise from a more complex bonding system than what simple models - frequently used in the literature - imply. 


\subsection{Classification Method of Pyrochlores}

\begin{figure}[h]
    \centering
    \resizebox{\textwidth}{!}{\includegraphics{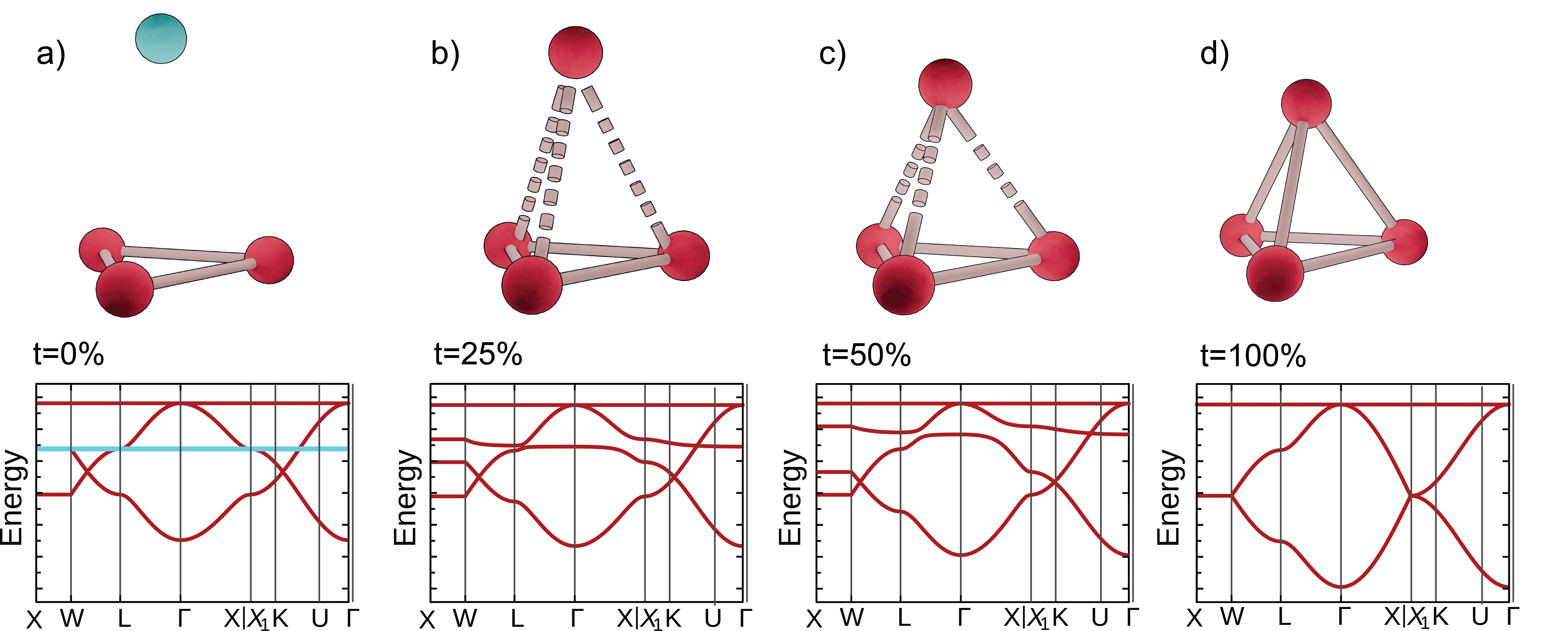}}
    \caption{Band Structure evolution from kagome to pyrochlore. As bonding is introduced between the kagome layer (red) and the isolated atom (blue), the band from the latter (shown in blue) interacts with the kagome bands, ultimately forming doubly degenerate pyrochlore flat bands. Here, t indicates the percentage bonding interaction between the kagome net and the isolated atom. }
    \label{kagome to pyrochlore}
\end{figure}

The pyrochlore lattice, as well as its band structure, can be understood starting from a kagome lattice. Fig.~\ref{kagome to pyrochlore} shows the evolution of both the lattice and the band structure as the kagome lattice evolves into a pyrochlore lattice. The pyrochlore network can be viewed as a kagome network where every second triangle is capped by the same atoms so that a 3D network of corner-sharing tetrahedra is formed. In a tight-binding model based on \textit{s} orbitals, this extra atom will initially add a non-interacting flat band (shown in blue in Fig.~\ref{kagome to pyrochlore} (a)). When the atom approaches the kagome layer, the bands mix, ultimately forming typical pyrochlore bands with a doubly degenerate flat band at the highest energy. The extent of this band mixing is influenced by the bonding interaction (t), which quantifies the percentage bonding between the kagome layer and the isolated atom.

\begin{figure}[H]
    \centering
    \resizebox{\textwidth}{!}{\includegraphics{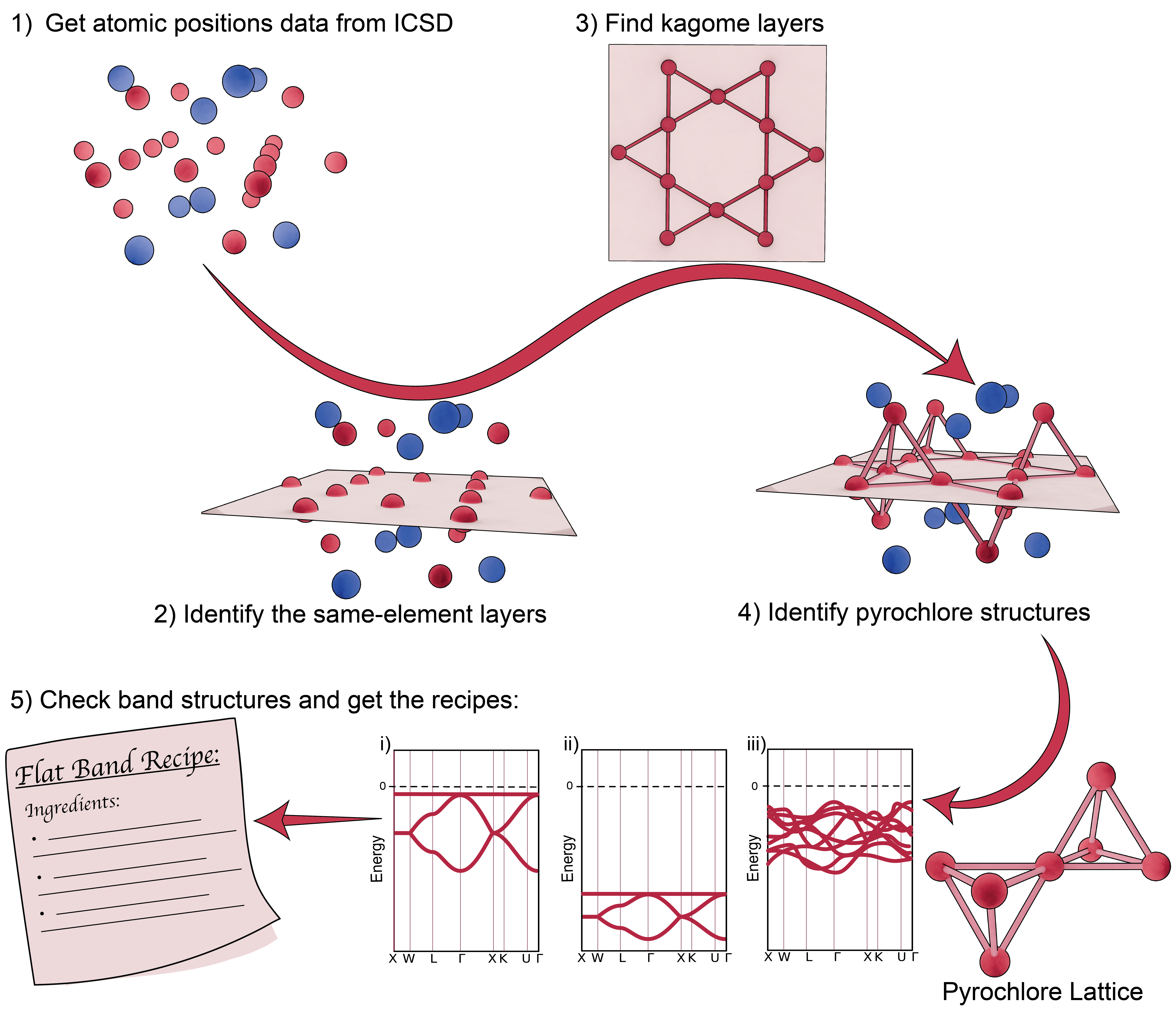}}
    \caption{Algorithm to detect pyrochlore sublattices and obtain recipe for flat bands: 1) Extract atomic positions from ICSD. 2) Scan each layer in the x-,y- and z-directions formed by the same elements. 3) Enforce geometric constraints to detect the same element kagome layer. 4) Check for the same element above and below and identify whether they form a pyrochlore sublattice by enforcing geometric constrains. 5) For all pyrochlore lattice containing materials, categorize the band structure into one of the three categories: (i) pyrochlore flat band near the Fermi level  ii) pyrochlore flat band at low energies, iii) no pyrochlore flat band observed. Identify common ingredients to deduce recipe for flat bands.}
    \label{pyrochlore}
\end{figure}

As the pyrochlore lattice consists of corner-sharing tetrahedrons, which have a kagome lattice (corner-sharing triangles) in their (111) plane (for cubic structures), we can take advantage of our previous work that identified all kagome materials in the ICSD to find pyrochlore lattices.\cite{Jovanovic2022}
Within this data, we searched for nearest neighbors from other layers that have the same atom type as the kagome layer. If those atoms form corner-sharing tetrahedrons with atoms in the kagome plane, the compound is classified as a pyrochlore compound. From this search, we identified 945 pyrochlore sublattice containing compounds in the ICSD.

We then analyzed the band structures of the identified pyrochlore compounds using Topological Quantum Chemistry and Materials Project\cite{Bradlyn2017,Jain2013, Vergniory2019, Vergniory2022,topologicalquantumchemistry, bilbao}. Our findings reveal that the band structures fall into three main categories: Fermi-level pyrochlore bands (17 compounds), pyrochlore bands at energies close to Fermi level (195 compounds), and no pyrochlore bands near the Fermi energy (750 compounds). A complete list of those classified compounds can be found in the supplemental file ``Pyrochlore Compounds List.pdf''. Note that the last category hosts by far the most compounds, which again highlights that the lattice type alone is not sufficient to exhibit the desired band structure. In the following sections, we will discuss several families of compounds with pyrochlore lattices and categorize their electronic structures accordingly. Importantly, we will discuss which chemical ingredients, beyond merely having a pyrochlore lattice in the structure, are necessary to exhibit pyrochlore bands in a given material.
A schematic of our materials search algorithm and band structure analysis is shown in Fig.~\ref{pyrochlore}.

\subsection{Cubic Laves Phases: AB$_2$}

\begin{figure}[H]
    \centering
    \resizebox{\textwidth}{!}{\includegraphics{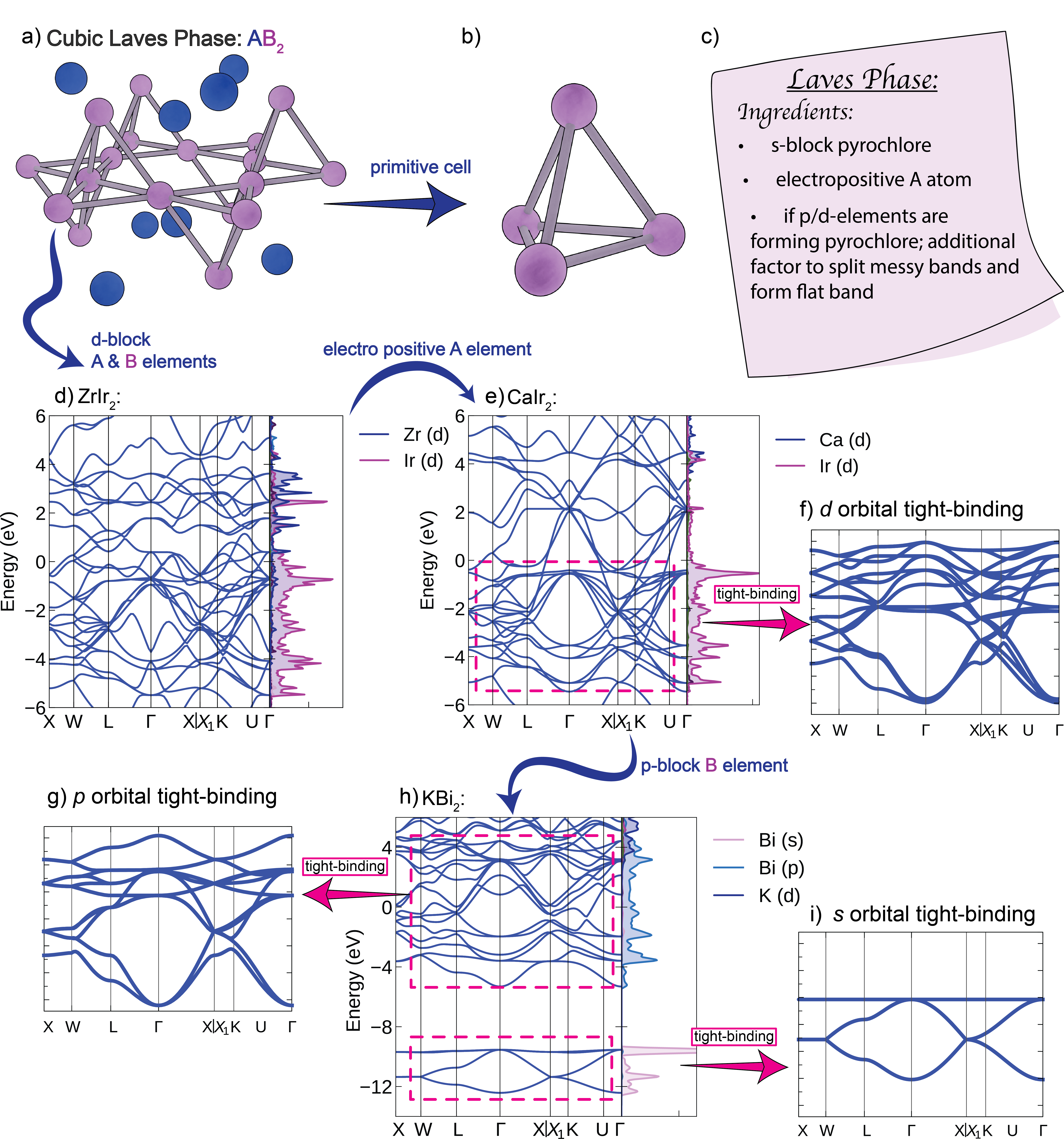}}
    \caption{Flat band recipe for the cubic Laves phase: a) Structure of the cubic Laves phase \ch{AB2}, where A atoms are shown in blue and B atoms in purple. The B atoms form the pyrochlore sublattice. b) Primitive cell of the pyrochlore sublattice. c) Recipe for Fermi-level flat bands in the cubic Laves phase d, e, h) Band structure evolution as A and B atoms are varied. Note that the pyrochlore flat band appears at lower energies (between -9 and -12 eV) in (h). f, g, i) Tight-binding models of the primitive cell shown in (b), with (i) \textit{s}-orbitals, (g) \textit{p}-orbitals, and (f) \textit{d}-orbitals.}
    \label{KBi2}
\end{figure}

One of the most important and simplest structure types that contains a pyrochlore lattice is the cubic Laves phase, shown in Fig.~\ref{KBi2}(a). Its general formula is AB$_2$ where B atoms form a pyrochlore pattern around A atoms\cite{Hoffmann_Laves}. We identified 250 cubic Laves phase materials as candidates for flat pyrochlore band compounds.  At first glance, the structure is a promising host for ideal pyrochlore bands because there is a great diversity of A and B atoms that can form the structure, from alkali metals to rare earths to $p$-elements (three examples are shown in Figs.~\ref{KBi2}(d), (e) and (h)). Thus, in principle, if the A atom is an electropositive element and the B element is an electronegative one, then the electron hopping, or bonding, is within the pyrochlore sublattice. Therefore, clean pyrochlore bands can be expected. 

However, as we will show now, in most cases the bands at the Fermi level do not represent the typical pyrochlore bands (see Figs.~\ref{KBi2}(d), (e) and (h)). In some cases, we found clean pyrochlore bands at low energies (-12 to -9 eV), but never at the Fermi level. An example of such a case is KBi$_2$, shown in Fig.~\ref{KBi2} (h). In KBi$_2$, it is reasonable to assume that potassium donates its electron to the Bi network and that thus the Bi pyrochlore bands should exist around the Fermi energy.  Elemental and orbital analysis of the bands reveals that indeed all bands close to the Fermi level come from Bi atoms. The problem is that Bi is a $p$-block element, but the textbook pyrochlore bands are derived from \textit{s} orbitals. Indeed, comparing tight-binding models of \textit{s} and \textit{p} orbitals on the pyrochlore lattice reveals very different band structures (Fig.~\ref{KBi2} (g,i)). Specifically, \textit{s} orbitals form perfect pyrochlore bands, which can be found at energies well below the Fermi level in KBi$_2$. But because the three \textit{p} orbitals are close in energy (degenerate if you do not consider the bonding), their bands mix and disperse. The resulting dispersive bands resemble those near the Fermi level in KBi$_2$. 
We can also understand why \textit{p} orbitals do not necessarily form clean pyrochlore bands if we go back to the orbital picture shown in Fig. \ref{intro_figure} (c). In contrast to the kagome lattice, where in the 2D system, all the \(p_z\) orbitals are orthogonal to \(p_x, p_y\),in the pyrochlore lattice, the \(p_z\) and \(p_x, p_y\) orbitals of the next-nearest neighbors are not necessarily orthogonal to each other. As a result, they mix and lead to a more complicated \(p\)-orbital band structure compared to the clean  \(s\)-orbital pyrochlore band structure. We can use group theory to identify which orbitals have this property, as we will do later.


The story is similar, but not identical, for the pyrochlore \textit{d}-bands. Fig.~\ref{KBi2}(f) shows a tight-binding model for five \textit{d} orbitals on a pyrochlore lattice. Similarly to the p bands, the bands interfere, causing changes in the typical band structure and dispersion in the flat band. However, because the overlap between the \textit{d} orbitals is smaller, the bands are less dispersed. Such bands can be realized in Laves phases with an electropositive A atom and a transition metal for B (Fig.~\ref{KBi2}(d,e)). Recently, a quasi-flat band was experimentally realized in \ch{CaNi2}\cite{Wakefield2023}; although the dispersion of this band is 0.5 eV, there are regions in the Brillouin zone where the dispersion is almost zero. This is an example of how \textit{d} orbitals on a pyrochlore lattice can lead to quasi-flat bands, despite their mixing.

One of the complicating factors of \textit{p} and \textit{d} orbitals is that they do not all align well for bond formation on this lattice, which decreases orbital overlap and bond strength. We will now use group theory to identify particular orbitals that are correctly aligned to maximize bonding. The point group of a pyrochlore structure is D$_{3d}$ (Table S1). This local symmetry group causes the \textit{p} and \textit{d} orbitals to split. The d$_{z^2}$ orbitals, for example, transform as A$_{1g}$, identical to \textit{s} orbitals, which is why we expect them to form the same kind of bands. All $p$ orbitals transform differently than $s$ orbitals under this point group, but \textit{p$_z$} orbitals, transforming as A$_{2u}$, will be allowed to form very similar bands if the $\sigma$ and $\pi$-bonding contributions are set equal in the tight-binding model. Wakefield et al. showed \cite{Wakefield2023} that \textit{d$_{z^2}$} orbitals can form the same bands as \textit{s} orbitals and that they can also be somewhat separated in Laves phases. Thus, in order to have a flat band one must (a) separate the orbitals and (b) use orbitals that have the right symmetry.

A recipe for pyrochlore bands in Laves phases is given in Fig.~\ref{KBi2} (c). If both lattice positions of the structure shown in Fig.~\ref{KBi2} (a) are occupied by \textit{d}-block elements, the band structure is very metallic as the example of \ch{ZrIr2} Fig.~\ref{KBi2} (d) shows. There are some similarities between the bands in the energy range of -4 to 0 eV and the \textit{d} orbital tight binding model shown in Fig.~\ref{KBi2} (f). If the non-pyrochlore atoms are replaced with an electropositive element, as shown with the example of \ch{CaIr2} in Fig.~\ref{KBi2} (e), the bands resemble those of the tight binding model even more closely. This is similarly true if a \textit{p}-block element is on the pyrochlore lattice and an electropositive element on the A sublattice, as shown in Fig.~\ref{KBi2} (h)  with the example of \ch{KBi2}. In this case, as mentioned above, the bands at the Fermi level resemble the bands that \textit{p} orbitals form on a pyrochlore level, and the textbook pyrochlore bands formed by the \textit{s} orbitals can be found at low energies. Thus, in order to have clean pyrochlore bands at the Fermi level one would need to have an \textit{s}-block element on the pyrochlore position, paired with a more electropositive A block element. However, such compounds do not exist, which is why the best strategy to find flat bands in Laves phases will be in the \textit{d}-block materials, even if those are not the textbook pyrochlore bands.

To conclude, cubic Laves phases can exhibit pyrochlore bands under certain conditions. These phases typically show well-defined pyrochlore bands at very low energies, originating primarily from \textit{s} orbitals. However, we are not aware of any cases where \textit{s} orbitals contribute significantly to bands at the Fermi level in Laves phases. Instead, the bands at the Fermi level are usually dominated by the \textit{p} or \textit{d} orbitals of the elements forming the pyrochlore lattice. Although these bands may follow the pyrochlore tight-binding model, the multiplicity of \textit{p} and \textit{d} orbitals tends to induce mixing and dispersion, reducing the flatness of the band. This effect is less pronounced for \textit{d} orbitals due to their lower overlap, which is why quasi-flat bands can still appear in Laves phases with transition metals on the pyrochlore lattices, as demonstrated by \ch{CaNi2}\cite{Wakefield2023}. Other Laves phases have also been investigated for their pyrochlore bands, such as YCo$_2$, which exhibits itinerant magnetism\cite{YCo2,YCo_2}, or ZrV$_2$, which shows a change in the superconducting transition temperature with Hf doping \cite{ZrV2,ZrV2_2}. In total, we identified 250 cubic Laves phases, the majority of which are \textit{d} orbital-based systems (see Supplementary Pyrochlore Compound List).  192 of these compounds have \textit{d}-elements on the B site and 172 of those exhibit pyrochlore-derived $d$-bands within the energy range of 2 eV to -2 eV, providing a significant pool of candidates for flat-band studies. Beyond \ch{CaNi2}, notable examples include \ch{CaIr2}, \ch{ZrIr2}, \ch{BaRh2}, and \ch{ScCo2}, \ch{ZrFe2}, \ch{LiPt2} and \ch{SrIr2} (band structures can be seen in Fig.S2).  42 compounds contain $p$-block elements, such as \ch{KBi2}, \ch{NdS2}, and \ch{MgIn2}, while only 16 compounds have \textit{s}-block elements, including \ch{SnMg2} and \ch{TaBe2}. In those latter cases, the \textit{s}-block elements are cations, thus not eligible for the formation of pyrochlore bands close to the Fermi level. Interestingly, no examples with f-block elements were identified in this dataset.

\subsection{Spinels: AB$_2$X$_4$}

\begin{figure}[H]
    \centering
    \resizebox{\textwidth}{!}{\includegraphics{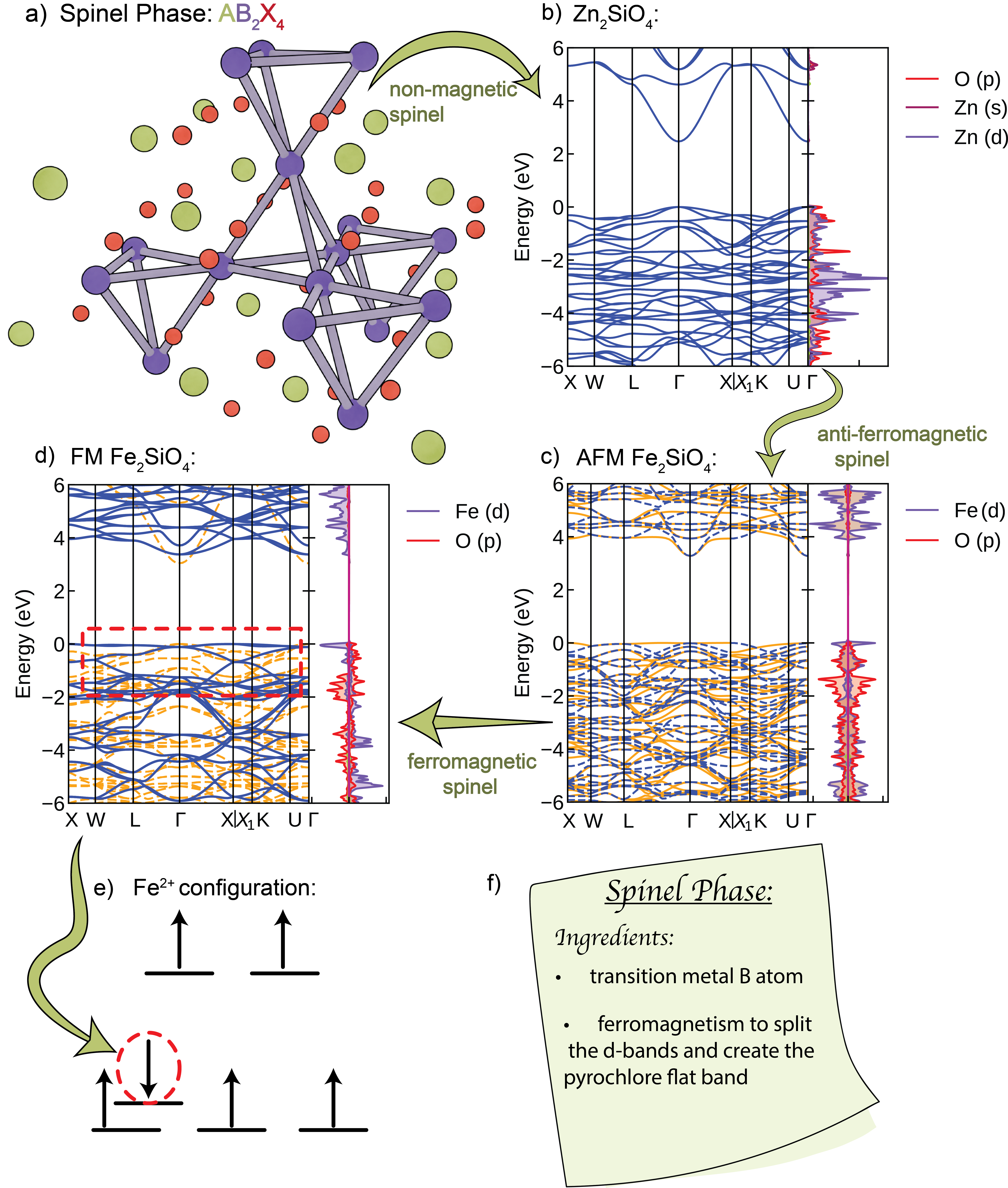}}
    \caption{Flat band recipe for spinels: a) Crystal structure of AB$_2$X$_4$ spinels (A: green,
B: purple, X:red) b) Non-magnetic band structure of \ch{Zn2SiO4} c) Anti-ferromagnetic band structure of \ch{Fe2SiO4} d) Ferromagnetic band structure of \ch{Fe2SiO4} e) Fe$^{2+}$:d$^6$ spin polarized high spin configuration f) Recipe of spinel pyrochlore flat bands. }
    \label{Fe2SiO4}
\end{figure}


Spinels are another prominent structure type that contains a pyrochlore sublattice. These compounds may exhibit magnetic order if they contain \textit{d}-block transition metals.  The spinel structure consists of two different metal ions, A$^{2+}$ and B$^{4+}$, and four chalcogen atoms (X$^{2-}$ - typically oxygen) (Fig. \ref{Fe2SiO4} (a)). In this structure, the B atoms form the pyrochlore sublattice, and they are most often transition metals. To observe flat bands in spinels, careful selection of elements is crucial. If the \textit{d} orbitals of the B atoms are close in energy to orbitals of either the A or X atoms, bonding will disrupt the pyrochlore bands. As was true for Laves phases, also here, quasi-flat bands can arise when \textit{d}-elements form the pyrochlore lattice. In the case of spinels, magnetism can act in favor to split the bands, as it can separate \textit{d}-manifold into spin-up and spin down.

This is most evident when considering ferromagnetic \ch{Fe2SiO4} as an example. The relevant absolute orbital energies are: Fe-\textit{d} = -11.66 eV, O-\textit{p} = -15.85 eV, and Si-\textit{s} = -14.89 eV. The energy separation between the Fe \textit{d}-states and other orbitals allows for the formation of pyrochlore \textit{d}-bands. Additionally, the \textit{d}-bands are split in energy, producing a distinct flat pyrochlore band at the Fermi level. Notably, this feature is only observed in spin-polarized calculations for ferromagnetic order (Fig.\ref{Fe2SiO4} (d)), suggesting that magnetic exchange, which splits the \textit{d}-bands into spin-up and spin-down states, is essential for the emergence of flat bands. Fe$^{2+}$ has six $d$ electrons, and under a octahedral crystal field with a high-spin configuration in a spin-polarized system, only one electron occupies the T$_g$ spin-down state. This results in the formation of the pyrochlore band (Fig.\ref{Fe2SiO4} (e)).

Although the thermodynamically stable structure of \ch{Fe2SiO4} is olivine, the spinel structure can be synthesized under high pressure\cite{Ringwood1966Mg2SiO4}. Experimentally, spinel \ch{Fe2SiO4} is found to be antiferromagnetic, differing from our calculations\cite{Yamanaka2001Fe2SiO4}. We emphasize the importance of magnetic splitting, as antiferromagnetic calculations reveal that in this case the pyrochlore bands are not isolated (Fig.\ref{Fe2SiO4} (c)). Even if the ground state is antiferromagnetic, a fully polarized state can often be achieved with an applied field, suggesting that the predicted flat bands in spinel \ch{Fe2SiO4} could be realized. In fact, we found that the energies of the AFM and FM states are very close, with the FM state being 150.82 meV more stable in our calculations. \ch{Co2SiO4}, which is also antiferromagnetic but can become fully polarized with a small applied field\cite{Lisboa2000}, although the authors of the paper do not specify whether they measured the olivine or spinel structure. In the spinel case, ferromagnetic calculations reveal a splitting in the $d$ bands, leading to the formation of the desired flat band (Fig. S4). Notably, Co$^{2+}$ has one hole in the T$_g$ spin-down state, which is why the pyrochlore band appears in the unoccupied states.

Additionally, the related spinel-type \ch{Co2GeO4} has been found to exhibit both antiferromagnetic and ferromagnetic interactions, as well as multiple metamagnetic transitions, indicating that a fully polarized state might be achievable\cite{HOSHI2007e448}. The ferromagnetic band structure shows similar magnetic flat bands at the Fermi level as in other spinels (Fig. S4). The ferromagnetic alignment in \ch{Fe2SiO4}, \ch{Co2SiO4}, and \ch{Co2GeO4} induces significant band splitting due to magnetic exchange, which supports the appearance of flat bands.

For comparison, we can examine a non-magnetic analog, \ch{Zn2SiO4} (Fig.\ref{Fe2SiO4} (b)). In \ch{Zn2SiO4}, the bands near the Fermi level include contributions from both Zn and O atoms, and without magnetism, the d bands remain unsplit. As a result, the band structure around the Fermi energy resembles the \textit{d} orbital tight-binding model without clear flat bands. Thus, we can summarize a recipe for flat bands in spinels which entails that we need a transition metal of the B-site, and electropositive elements on the A-site and ferromagnetism to split bands into spin up and spin down channels, which then can result in more isolated \textit{d} orbitals.

To summarize, our investigation highlights several spinel structures that are exhibiting magnetic flat bands at the Fermi level due to ferromagnetic band splitting facilitated by magnetic exchange. As discussed earlier, \ch{Fe2SiO4}, \ch{Co2SiO4}, and \ch{Co2GeO4} serve as prime examples, where ferromagnetic alignment enables significant band splitting that supports the emergence of flat bands. Additionally, other spinels, including \ch{Fe2GeO4}, \ch{Cr2ZnO4}, \ch{V2AlO4}, \ch{V2MgO4}, \ch{V2LiO4}, \ch{V2CdO4}, \ch{V2FeO4}, \ch{V2ZnO4}, and \ch{V2MnO4} (Fig.S4), exhibit similar behavior, emphasizing the diverse chemical and structural environments in which magnetic flat bands can arise. Together, these examples highlight these spinels as promising candidates for exploring the relationship between magnetism and pyrochlore flat band phenomena. While the magnetism-induced band splitting clearly helps to separate the bands, those flat bands are still not fully isolated. This brings us to the next class of materials, where we find the cleanest pyrochlore bands among all known materials.

\subsection{Pyrochlore Oxides: A$_2$B$_2$O$_6$O’}

\begin{figure}[H]
    \centering
    \resizebox{0.8\textwidth}{!}{\includegraphics{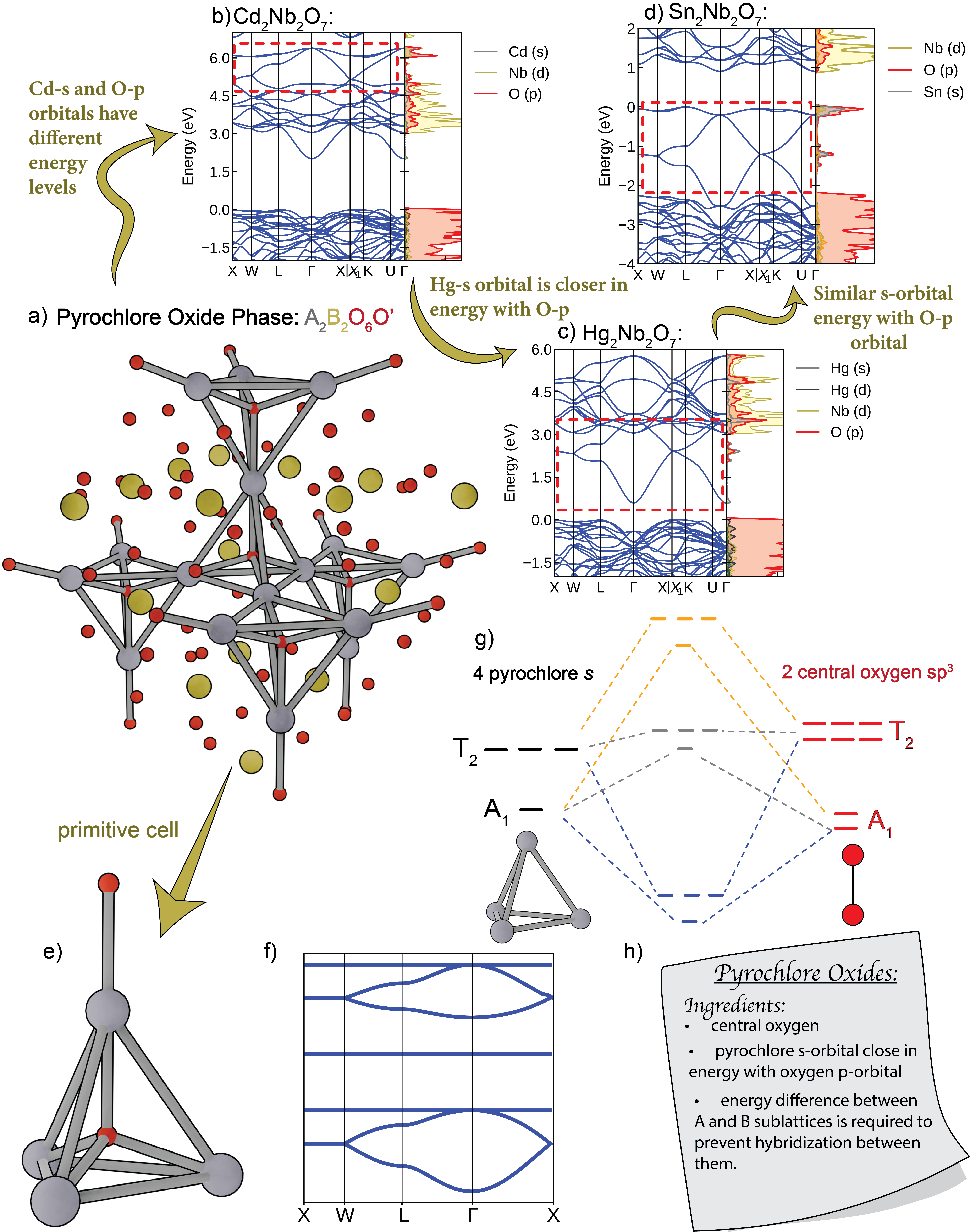}}
    \caption{Flat band recipe for pyrochlore oxide phase: a) The structure pyrochlore oxides A$_2$B$_2$O$_6$O' (A atoms are illustrated as gray color, B atoms are in yellow  and all the oxygen atoms are in red color). Both A and B form a pyrochlore sublattice but the central oxygen is in A sublattice. b) The band structure of \ch{Cd2Nb2O7} c) Making the energy level of pyrochlore $s$ orbital closer with oxygen by changing pyrochlore atom from Cd to Hg. d) Changing the pyrochlore lattice with $p$-block element yields flat band at the Fermi level. e) Primitive cell of pyrochlore sublattice with the central oxygen. f) Tight-binding model of pyrochlore sublattice with central oxygen \textit{sp$^3$} orbitals. Models shows that bonding and anti-bonding orbitals are forming the clean pyrochlore shape and fourthly degenerate non-bonding orbitals lies between them. g) MO-diagram of primitive cell. Anti-bonding (orange), non-bonding (gray) and bonding (blue) matches with the tight-binding model and proves that the flat band stems form lattice frustration. i) Recipe of pyrochlore oxide flat bands.} 
    \label{A2B2O7}
\end{figure}

Compounds in the A$_2$B$_2$O$_6$O’ structure type are the name-giving compounds to the pyrochlore structure. This structure consists of two interpenetrating pyrochlore lattices formed by A and B elements, respectively. The tetrahedra of both pyrochlore lattices are bridged by oxygens. Moreover, the A sublattice also has an additional oxygen (O') at the center of its tetrahedra. Therefore, the structure is much more complex compared to the Laves phase. Famous examples of this family are the pyrochlore iridates,  such as \ch{Pr2Ir2O7}\cite{Ueda2017}, which were early predictions for Weyl semimetals\cite{Ueda2018}. Intuitively, one would think that the complexity would complicate the band structure. Indeed, in earlier work \cite{Jovanovic2022, Klemenz2020} on other lattice types, we showed that orbitals of other atoms in close proximity to the lattice of interest often disrupt the desired band structure.  In contrast, we find here that some pyrochlore oxides yield very isolated pyrochlore bands, including a flat band at the Fermi level, as illustrated by \ch{Sn2Nb2O7} in Fig.~\ref{A2B2O7}(d). 

We now seek to understand this phenomenon. We first note that the orbital projections of these Fermi-level pyrochlore bands have contributions from both the \textit{s} orbitals of the A pyrochlore element (here Sn), and the \textit{sp$^3$} hybrid orbitals of the central oxygen within the A tetrahedra. In the case of Laves phases, the $s$-bands were well below the Fermi level, but here $s$-bands are at the Fermi level. However, the A element's \textit{s}-bands are mixed with states from the central oxygen atom. In addition, \textit{s} orbital contributions from the A element can also be found at lower energies, around -8 eV as can be seen in Fig.S12. This implies that the \textit{s} orbitals split into two parts, possibly as a result of the bonding with the central oxygen. 

We test this hypothesis by constructing a tight-binding model that includes both the A element's \textit{s} orbitals, and the orbitals from the central oxygen. A conventional pyrochlore lattice has four atoms per unit cell, which is why the typical pyrochlore band structure with \textit{s} orbitals has four bands. Due to the close proximity of the central oxygen atom to the pyrochlore lattice, a model that accurately describes the oxide pyrochlores requires more atoms, and thus will have more bands. A repeating unit of an A-atom tetrahedra includes the two central oxygen atoms, as shown in Fig.~\ref{A2B2O7}(e). The tight-binding band structure of this primitive cell can be seen in Fig.~\ref{A2B2O7}(f). It is evident that bonding of the pyrochlore \textit{s} orbitals with the central oxygen splits the pyrochlore \textit{s}-bands.

The molecular orbital (MO) diagram of the model helps explain how the bonding results in the corresponding band structure (see Fig.\ref{A2B2O7}(g)). The MO diagram reveals anti-bonding, non-bonding, and bonding MOs, each of which will form bands. This is mirrored in the tight-binding band structure (Fig.\ref{A2B2O7}(f)), which shows that the bonding and anti-bonding MOs form typical pyrochlore bands and the non-bonding MOs will form a non-interacting trivial flat band.  Thus, the flat band at the Fermi level comes from anti-bonding orbitals of the O$_2$A$_4$ tetrahedra. Thus, this flat band arises due to destructive interference in the bonding network, making it a ``desired'' flat band. We further note that the oxygen bridging the tetrahedra atoms do not contribute to the relevant bands, justifying the model. The orbital contribution of the bridge-oxygens lies between -4 to -7 eV. The separation of the contribution from different oxygen atoms can be understood by the orbital geometries, as the central oxygen \textit{sp$^3$} orbitals align perfectly with the pyrochlore tetrahedra, so the A-atom \textit{s} orbitals easily bond with them but not with the non-aligned bridge \textit{p} orbitals. This model can be used to universally describe a whole fleet of A$_2$B$_2$O$_6$O’, not only \ch{Sn2Nb2O7}, which we used as an example here. In Fig.S3 we show the band structures of \ch{Sn2Ta2O7}, \ch{Tl2Nb2O7}, \ch{Pb2Ta2O7}, \ch{Pb2Nb2O7}, \ch{Tl2Ru2O7}, which are all related and exhibit pyrochlore flat bands at or around the Fermi level.

The common characteristics of these compounds lie in the energy levels of the elements in both the A and B sublattices. According to molecular orbital (MO) theory, bonding interactions can occur only if: 1) the energies of the states are similar and 2) the symmetries of the orbitals allow an interaction. In this case, a flat band at the Fermi level emerges only when the \textit{s} orbital energy of the A sublattice closely matches the \textit{p} orbital energy of oxygen. Therefore, it is essential to select pyrochlore atoms with \textit{s}-orbital energy levels close to those of the oxygen \textit{p} orbitals to facilitate bonding. The absolute energies of the corresponding orbitals of the elements are as follows: the oxygen \textit{p} orbitals are positioned at -15.85 eV, while the \textit{s} orbitals of Sn, Pb, and Tl are at -14.56 eV, -15.12 eV, and -13.14 eV, respectively. Those energies are close enough to facilitate bonding. Notably, the cases of \ch{Cd2Nb2O7} and \ch{Hg2Nb2O7} demonstrate that when the energy of the A sublattice significantly deviates from the oxygen \textit{p} orbital energy level, the pyrochlore bands shift above the Fermi level (Fig.\ref{A2B2O7}(b,c)). Additionally, the energy level of the B sublattice should differ sufficiently to avoid interference with the bonding between the A-site pyrochlore and the central oxygen. Consequently, Nb and Ta, with absolute \textit{d} orbital energies of -8.86 eV and -7.58 eV, respectively, are ideal candidates for this role. The band structures of these combinations reveal that, as the energy gap with the oxygen \textit{p} orbital narrows, the likelihood of a flat band appearing at the Fermi level increases, in line with MO theory.

To summarize, we find here that the cleanest pyrochlore bands are found in compounds where the typically used tight-binding model for pyrochlore bands does not accurately describe the system. Instead, we need to take orbitals from the central oxygen atom, which are bonding with the pyrochlore \textit{s} orbitals, into account as well. This results in a band structure that features a set of bonding, non-bonding and antibonding bands, where the non-bonding orbitals form a trivial (or ``boring'') flat band, whereas the bonding and antibonding orbitals form bands that resemble the textbook pyrochlore bands. The splitting induced by bonding lifts those \textit{s}-type bands to the Fermi level, which explains why pyrochlore oxides allow for such pyrochlore bands at the Fermi level, in contrast to other structures where such \textit{s}-bands will always be far below the Fermi level.

Figure \ref{A2B2O7} presents our framework for designing pyrochlore oxides with flat bands at the Fermi level, combining structural, chemical, and electronic insights. Here, we emphasize how the geometry of the pyrochlore lattice and the placement of central oxygen atoms within the A- or B-sublattices play a crucial role in determining the flat band characteristics. The sublattice containing the central oxygen is responsible for the formation of the flat band, and its influence on the electronic structure is evident. By tuning the energy levels of the pyrochlore atoms, such as replacing Cd with Hg, we align them closer to the oxygen states, enhancing the flat band formation. The structural basis of this phenomenon can be depicted through the primitive cell of the pyrochlore lattice, and its electronic characteristics can be understood using a tight-binding approach, validated by molecular orbital diagrams. Our MO analysis (Fig.S5-13) reveals that bonding and anti-bonding states contribute to the formation frustration-induced flat bands. Overall, a step-by-step recipe can be derived, linking structural design and electronic features, to guide the creation of ideal pyrochlore oxide phases.

Similar characteristics can also be found in compounds with \textit{p}-block elements in the tetrahedra. For example, in \ch{Hg2Cu2F6S}, \ch{Hg2Ni2F6S}, and \ch{Hg2Mn2F6S}, the central atom is sulfur rather than oxygen and it is located at the center of the Hg tetrahedra. Despite this difference, the band structures of these compounds follow the same fundamental rules. As shown in Fig. S3, clean pyrochlore bands are primarily formed by pyrochlore \textit{s} orbitals and sulfur \textit{p} orbitals. Furthermore, the band structures of \ch{Hg2Ni2F6O} and \ch{Hg2Zn2F6O} clarify that bridging oxygens are not essential for the formation of the clean pyrochlore bands (Fig. S3). In these structures, fluorine takes the role of the bridge-oxygen instead of oxygen, yet the characteristic electronic structure remains unchanged, highlighting the dominant role of pyrochlore \textit{s} and oxygen \textit{p} orbitals in shaping the band dispersion.


Note that some of those compounds, while listed in the ICSD, may not be stable, as oxygen vacancies and distortions are common in A$_2$B$_2$O$_6$O’ compounds\cite{Marchetti2020, Aiura2017}. Therefore, we will now proceed to understand how such vacancies affect the electronic structure and particularly the flat bands.

\paragraph{Oxygen Deficient A$_2$B$_2$O$_6$O$_x$’ phases}

\begin{figure}[H]
    \centering
    \resizebox{0.88\textwidth}{!}{\includegraphics{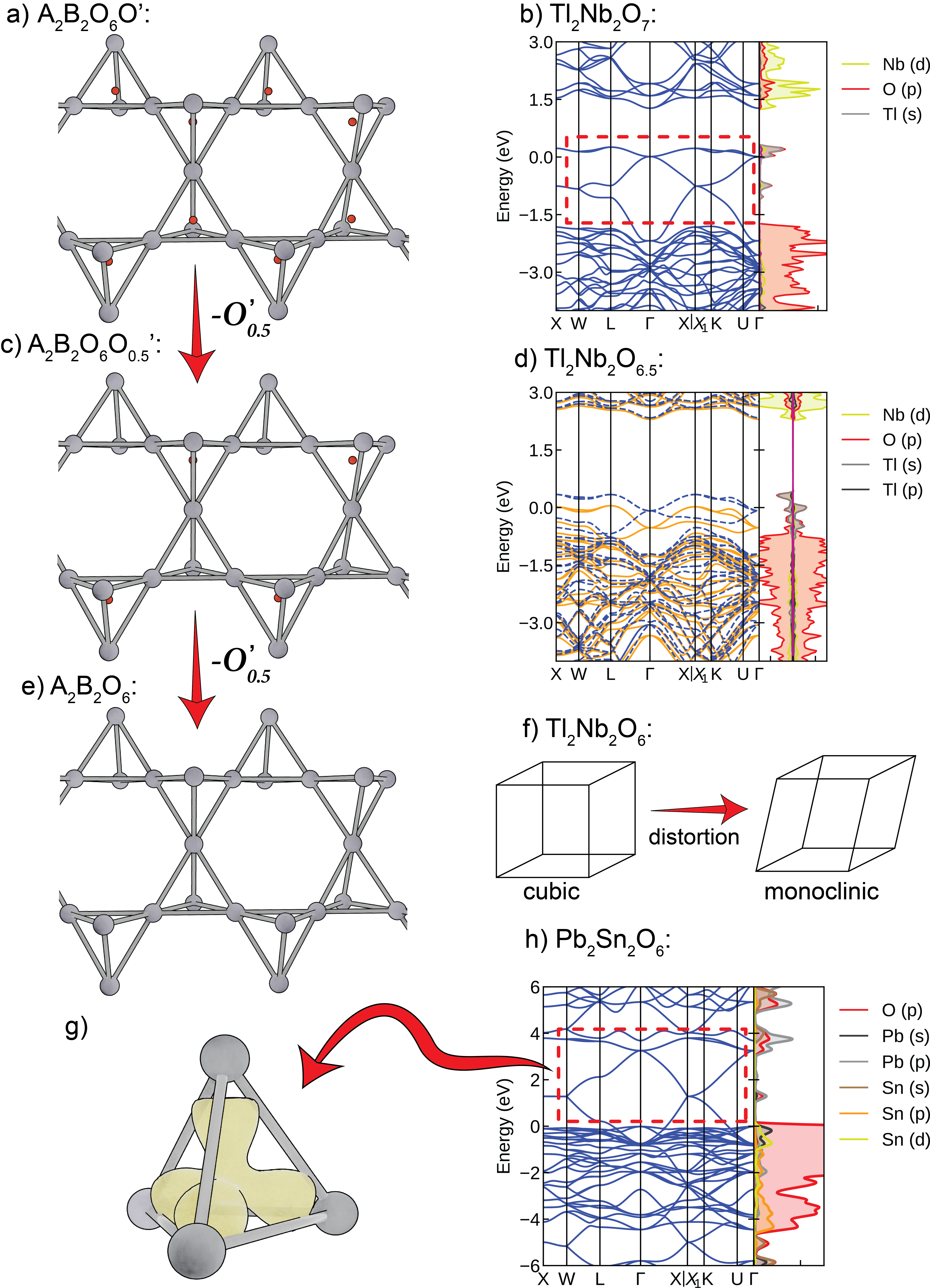}}
    \caption{Oxygen deficient pyrochlore oxide phases: a,c,e) The structures of \ch{A2B2O6Ox} b,d) Band structures of materials examples for each case shown on (a,c). f) No central oxygen may lead to structural distortion. g) Partial charge density calculation on \ch{Pb2Sn2O6} h) A stereoactive lone pair in an oxygen deficient pyrochlore oxide phase prevents structural distortion and splits the \textit{p}-bands, leading to flat band formation.} 
    \label{Pb2Sn2O6}
\end{figure}

As derived above, achieving an \textit{s} orbital pyrochlore flat band at the Fermi level relies heavily on the role of the central oxygen atom, since the bonding interaction elevates the \textit{s}-bands to the Fermi energy. However, we found that A$_2$B$_2$O$_6$ compounds, which lack the central oxygen (see Fig.\ref{Pb2Sn2O6}(a,c,e)), may also exhibit such bands around the Fermi level, raising questions about the accuracy of our initial assumption. An example of such is \ch{Pb2Sn2O6} (Fig.\ref{Pb2Sn2O6}(h)), which exhibits clean pyrochlore bands around the Fermi level. An investigation of orbital contributions reveals that, in this case, the pyrochlore bands do not stem from \textit{s} orbitals but rather from the \textit{p} orbitals of the A element (Fig.\ref{Pb2Sn2O6}(h)). This suggests that something else within the sublattice must act comparably to the central oxygen in \ch{A2B2O7} phases, causing the \textit{p} orbitals to split into two groups: one set forming the pyrochlore bands, and the other set remaining separate. A previous study showed that Pb$_2$Sn$_2$O$_6$ exhibits a stereoactive electron lone pair, as evidenced via electron localization function (ELF) calculations \cite{Seshadri2006}. Stereoactive lone pairs do not localize exactly at the center of the tetrahedra but towards the centers of the triangles. This localization creates a crystal field environment that splits the \textit{p}-bands and creates pyrochlore motif. Examining the charge density of these pyrochlore bands (Fig.\ref{Pb2Sn2O6}(g)) reveals that the \textit{p} orbitals are oriented toward a central point, almost as if an atom were present there, aligning in a way that suggests bonding, similar to what we saw in A$_2$B$_2$O$_6$O' phases.

Additionally, there are compounds in which some tetrahedra contain a central oxygen while others are empty. For instance, in \ch{Tl2Nb2O7} (Fig.\ref{Pb2Sn2O6}(b)), well-defined pyrochlore bands emerge near the Fermi level when all tetrahedra are occupied by oxygen. However, introducing oxygen vacancies disrupts these bands. In the case of \ch{Tl2Nb2O6.5} (Fig.\ref{Pb2Sn2O6}(d)), where every other tetrahedron lacks the central oxygen\cite{Shoemaker2011}, centrosymmetry is broken due to the displacement of Tl atoms from their centrosymmetric positions\cite{Zhang2019}. This changes the symmetry of the lattice from a cubic centrosymmetric space group Fd-3m (SG 227) to a non-centrosymmetric cubic space group F-43m (SG 216). The loss of inversion symmetry significantly alters the band structure by allowing asymmetric band splitting and removing degeneracies that would otherwise be protected by inversion symmetry. In the extreme case of \ch{Tl2Nb2O6} (Fig.\ref{Pb2Sn2O6}(f)), the structure is distorted due to the absence of stabilizing lone pairs in the tetrahedra voids. Generally, the loss of central oxygen is detrimental to pyrochlore bands unless the structure includes a stereochemically active lone pair.

\subsection{Other Pyrochlore Materials}

Beyond the three main types of pyrochlore materials - Laves phases, pyrochlore oxides, and spinels — there are many other structures that include pyrochlore sublattices. These structures are not limited to the canonical SG-227, which is typically associated with the pyrochlore lattice, but also occur in space groups such as SG-9, 44, 46, 70, 74, 96, 141, 166, 194, 203, and 216. In our search we could not identify any materials with clean pyrochlore bands in those materials. We did however find that the trend observed for oxides above - namely that small \textit{p}-group atoms within the tetrahedra can form bonding interactions that lift characteristic bands to the Fermi level - seems to be a universal effect  ( see section ``Other Phases'' in the SI). As discussed in detail in the SI, we find the same phenomenon in distorted pyrochlores, however in this case the affected bands do not have the typical pyrochlore band shape.

\section{Potential Future Experimental Realization}

Although many of the compounds discussed here have frustrated flat bands close to the Fermi level, many may be hard to realize experimentally. On paper, a perfect flat band can exist, but in real life, nature often finds a way to avoid it, resulting in experiments inconsistent with the DFT predictions. A very flat band at the Fermi level usually indicates an instability in the structure and it cannot be ruled out that the real structures have slight distortions that were previously missed in experiments, or that real materials contain too many defects to validate the DFT predictions. In our search, the cleanest flat bands appeared exclusively in pyrochlore oxides. Figure S3 shows the compounds with clean flat bands that we found, which are not already shown in the main text. However, these compounds, especially the pyrochlore oxides, often struggle with non-stoichiometry, which can affect the band structure. For example, Pb$_2$Sb$_2$O$_7$, is listed in the ICSD and has a very flat band exactly at the Fermi level (Fig S3(h)). In fact, Pb$_2$Sb$_2$O$_7$, or its oxygen-deficient variants, have been well known since ancient times, when it was used as a yellow pigment: Naples Yellow\cite{WikipediaNaplesYellow, NaplesYellow}. The yellow color, of course, implies a large band gap and indicates that the ICSD-listed structure is very likely incorrect. This has been noticed before, and structural rearrangements and defects have been postulated to be the cause of the color \cite{Marchetti2020}. Oxygen-deficient \ch{Pb2Sn2O6} (Fig. \ref{Pb2Sn2O6} (i)) is a good photocatalyst, which implies the presence of a significant band gap,\cite{Wang2008Pb2Sn2O6} disagreeing with the DFT prediction of an ideal crystal. However, this study was based on nanocrystals, which often have more defects than single crystals, and does not rule out the possible synthesis of stoichiometric single crystals. Similarly, a discrepancy between the computed and measured band structure has been highlighted in \ch{Sn2Nb2O7} (Fig.\ref{A2B2O7}(e), which was attributed to off-stichometry in the samples\cite{Aiura2017}. This does not rule out the possibility that better crystals could be grown, possibly using fluxes or more advanced methods. 

We also found several materials, which on first glance seem feasible. An example is \ch{Tl2Nb2O7} (Fig.S3(d)). Although its oxygen-deficient cousin \ch{Tl2Nb2O6} has been mainly studied, it has been reported that the solid solution of \ch{Tl2Nb2O_{6+x}} can be synthesized up to x=1.\cite{Fourquet1995Tl2Nb2O6} No experimental papers on such crystals are available to the best knowledge of the authors. The compounds \ch{A2B2O7} with  (A = Sn, Pb; B = Nb, Ta (see Fig.S3), have all been synthesized in thin film form,\cite{Fujita2022TrendsBandgap} but we are unaware of any single crystal growth thus far. Of these, \ch{Sn2Ta2O7} has previously been highlighted as a good flat band material.\cite{flatbandcatalogue}  Interestingly, it was found experimentally via photoemission experiments that the emission spectra agree with DFT-predicted band gaps and quasi-flat bands\cite{Fujita2022TrendsBandgap}. The authors of the same study also concluded that the Pb-based compounds have a flatter valence band compared to that of the Sn-based ones. Another interesting family to look at is \ch{Tl2Ru2O_{7-x}} (Fig. S3(e)). Although those flat bands are not isolated, they can be found close to the Fermi level. Depending on the value of x, such samples have been found to be metallic, have metal-to-insulator transitions, show spin glass behavior, or even evidence of spin-Haldane chains\cite{Takeda1998,Lee2006}.

Another future avenue is to pursue compounds with other \textit{p}-elements than oxygen within the tetrahedra. However, the ones we found in the database and  show in Fig. S3 all show the pyrochlore bands above the Fermi level. It might be useful to design new compounds based on the orbital energies of the individual elements, such that they match to form the most efficient bonding.

Of course, Laves phases remain worth studying, especially when \textit{d} orbital based, as discussed above. Although these flat bands are never isolated, they do appear close to the Fermi level and have been suggested to be the cause of superconductivity that is also often found in those compounds\cite{Tian2023,Unconventional_superconductivity, ZrV2}. Besides the already studied \ch(CaNi2) \cite{Wakefield2023}, we highlight \ch{CaIr2}, \ch{ZrIr2}, \ch{BaRh2}, \ch{ZrFe2}, \ch{LiPt2}, \ch{SrIr2} and \ch{ScCo2} (Fig.S2) as worthy candidates for further study.

Finally, when it comes to spinels, we note that some might be particularly relevant if found to be ferromagnetic. For instance, \ch{Fe2GeO4}, \ch{Cr2ZnO4}, and \ch{V2ZnO4}, which exhibit magnetic flat bands in our theoretical analysis, remain promising candidates for experimental verification (see Fig. S4 for band structures). However, the synthesis of these spinels often requires high-pressure conditions, and only limited experimental studies have been conducted on them to date. Further investigation of these compounds, both computational and experimental, could provide valuable insights into the connection between magnetism and flat band phenomena in spinel systems.

\section{Conclusion and Outlook}

Flat bands arising from geometric frustration are desired because they give rise to strong electronic correlation associated with exotic properties. Achieving flat band compounds in bulk crystals can be difficult as a result of interference from other bands. Using pyrochlores as a basis, we here lay out ways to achieve flat bands at the Fermi level. First of all, we highlight that simply having a pyrochlore lattice in a crystal structure does not guarantee the desired pyrochlore bands, including the flat band. Maybe unsurprisingly, we find that it is desirable to have bands as isolated as possible and orbital manifolds that are close in energy will be unfavorable. More surprisingly, we find that bonding with atoms that are inside the pyrochlore tetrahedra is beneficial for creating flat bands as such bonding carries \textit{s}-type bands to the Fermi level. Symmetry breaking induced by lone pairs can be another avenue to separate bands. In addition, we have shown how magnetic order may create flat bands at the Fermi level. We discussed the most promising material candidates and experimental challenges. We hope that, in the future, this theoretical work, especially the new insight about the formation of molecular orbitals and how it can result in flat bands at the Fermi level, can be a guide for the experimental synthesis of 3D flat band materials.  For example, it might be beneficial to consider lattice networks that contain more than just a single element for the typical minimal models. Future models should also consider how the formation of molecular orbitals with neighboring elements can be helpful to generate flat bands. In this case, it is important to consider the orbital energies of the different components so that bonding can be maximized.

\section{Methods}

\paragraph{DFT}

To identify candidate materials, we analyzed band structures of pyrochlore materials from the Materials Project \cite{Jain2013} and the Topological Materials Database \cite{Bradlyn2017, Vergniory2019,Vergniory2022, topologicalquantumchemistry, bilbao}.

Density Functional Theory (DFT) calculations were carried out using the Vienna Ab initio Simulation Package (VASP) 5.4.4\cite{VASP_1, VASP_2}. For these calculations, we employed the Perdew-Burke-Ernzerhof (PBE) functional\cite{PBE} for exchange-correlation and the recommended Projector Augmented Wave (PAW) potentials for all elements\cite{PAW_1, PAW_2}. Calculations were performed on experimental geometries reported in the Inorganic Crystal Structure Database\cite{rudorff1943kristallstruktur}, without additional structural relaxation.

To visualize the band structures at specific points, partial charge density (PARCHG)\cite{vasp_band_decomposed} and electron localization function (ELF)\cite{Silvi1994} calculations were performed. Band structure plots were generated using the Sumo package\cite{Ganose2018}, with some calculations directly obtained via the Materials Project API\cite{Ong_2015}.

\paragraph{Tight-binding}

The Hamiltonian for simple pyrochlore systems was derived manually (section ``Hamiltonian'' in the SI). To explore more complex systems, such as those involving additional orbitals, central atoms, or modified lattice structures, we utilized both PythTB \cite{coh2022pythtb} and Tight-Binding Studio \cite{Tb-Studio}. The results from these tight-binding models were cross-validated with our group-theory analysis to better understand the chemical origins of the bonding interactions.

\begin{acknowledgement}
 We thank Scott B. Lee, Grigorii Skorupskii and Hyosik Kang for helpful discussions. This project was supported by NSF, grant OAC-2118310, and by NSF through the Princeton Center for Complex Materials, a Materials Research Science and Engineering Center DMR-2011750.
 J.C. acknowledges support from the National Science Foundation under Grant No. DMR-1942447, from the Alfred P. Sloan Foundation through a Sloan Research Fellowship, and from the Flatiron Institute, a division of the Simons Foundation.

\end{acknowledgement}

\bibliography{Library}

\end{document}